# Supporting interoperability of collaborative networks through engineering of a service-based Mediation Information System (MISE 2.0)

Frederick Benaben[1], Wenxin Mu[1], Nicolas Boissel-Dallier[1&2], Anne-Marie Barthe-Delanoë[1], Sarah Zribi[1&2], Herve Pingaud[3]

[1]*Centre Génie Industriel, Mines Albi-Carmaux – Université de Toulouse, Albi, France*

[2]*Linagora, Toulouse, France*

[3]*Centre Universitaire Jean-François Champollion, Albi, France*

{first.last}@mines-albi.fr; szribi@linagora.com ; herve.pingaud@univ-jfc.fr

# Supporting interoperability of collaborative networks through engineering of a service-based Mediation Information System (MISE 2.0)

The Mediation Information System Engineering project is currently finishing its second iteration (MISE 2.0). The main objective of this scientific project is to provide any emerging collaborative situation with methods and tools to deploy a Mediation Information System (MIS). MISE 2.0 aims at defining and designing a service-based platform, dedicated to initiating and supporting the interoperability of collaborative situations among potential partners. This MISE 2.0 platform implements a model-driven engineering approach to the design of a service-oriented MIS dedicated to supporting the collaborative situation. This approach is structured in three layers, each providing their own key innovative points: (i) the gathering of individual and collaborative knowledge to provide appropriate collaborative business behaviour (key point: knowledge management, including semantics, exploitation and capitalization), (ii) deployment of a mediation information system able to computerize the previously deduced collaborative processes (key point: the automatic generation of collaborative workflows, including connection with existing devices or services) (iii) the management of the agility of the obtained collaborative network of organizations (key point: supervision of collaborative situations and relevant exploitation of the gathered data). MISE covers business issues (through BPM), technical issues (through an SOA) and agility issues of collaborative situations (through EDA).

**Introduction**

Ensuring the link between business considerations and IT is probably the ultimate goal of an IS. MISE provides an approach to support this goal in a collaborative context. Furthermore, MISE surrounds this goal with two correlated goals (one upstream and one downstream): (i) the emergence of collaborative business process and (ii) the management of the agility of the collaborative IS.

Enterprises (of any kind) working in today's economic environment are deeply dependent on their ability to take part in collaborative activities [1]. Consequently, an important requirement is for them to be able to take part in emerging, potentially opportunistic, collaborative networks [2]. It is noticeable that *Organization Integration* and *Organization Interoperability* both concern the same issue (facilitating collaboration of organizations). However, *Integration* has a strong organizational dimension, while *Interoperability* is more of a technical nature. This article is more dedicated to *Interoperability,* in that the main goal of the presented research works is to support collaborative situations through computerized information systems. The concept of *Interoperability* has been defined in [3] as *the ability of two or more systems or components to exchange information and to use the information that has been exchanged*. From this initial definition (made almost 25 years ago), several complementary visions have been provided, such as in [4] by the InterOp Network of Excellence (NoE) as "the ability of a system or a product to work with other systems or products without special effort from the customer or user". Consequently, interoperability of organizations appears to be a major issue to succeed in building emerging enterprise networks "on the fly" [5]. Therefore, enterprises need to adopt the required interoperability functions: exchange of information, coordination of functions and orchestration of processes. Furthermore, inside these organizations, Information Systems (IS) and computerized systems are assuming the role of interfaces (external and internal exchanges), functional engines (driving processes and business activities [6]) and data providers (creating a drastically increasing amount of measurements, data and reports from devices, software and reporting tools). Thus, IS must be able to support the previously listed interoperability functions. The issue is to ensure that the partners' ISs will be able to work altogether (thanks to these interoperability functions) to constitute a coherent and homogeneous IS set (the IS of the collaborative network). Providing organizations with methods, tools and platforms able to ensure these interoperability functions makes good sense.

The MISE project (Mediation Information System Engineering) was launched in 2004 and is dedicated to providing an approach (and the associated tools) for Mediation Information System (MIS[1]) design, following the mediation principle as described in [7]. The overall objective is to meet both the expectations regarding collaborative situations and the preponderant role of the IS. The approach aims at defining a mediation system able to connect the whole set of partners' ISs in a way that is (i) coherent with the business objectives of the network (effective) and (ii) easy and fast to deploy (efficient). Furthermore, the MIS thus obtained should ensure the interoperability functions (translation of data, sharing of services and orchestration of workflows) in an agile manner. In reality, collaborations are very unstable situations requiring adaptation: contexts can change (new opportunities, modification of objectives, etc.), networks of partners can change (withdrawal or arrival of partner, lack of resources, etc.) or dysfunction during the collaborative behaviour can occur (even if context and partners are still the same, something may not happen as expected). Therefore, the MIS should remain well adapted to the potentially changing needs of the collaboration. The general approach of the MISE project is based on three levels: (i) *business level*: definition of the appropriate collaborative behaviour that fits with the business issues of the network, (ii) *technical level*: design of the SOA[2] system and deployment on an ESB[3] so that it can assume the role of mediator between all partners' ISs, and (iii) *agility level*: management of evolutions and changes required for the MIS.

---

[1] This article uses the acronym MIS for *Mediation Information System* even if it has also been known as the official acronym for *Management Information System* for years.

[2] Service Oriented Architecture is a paradigm for information systems

Figure 1. The MISE structure.

Two iterations of the MISE project have already been performed. MISE 1.0 is presented in [8], [9] and [10]. This article aims to present a complete overview of MISE 2.0 and how this version intends to support collaborative networks in the Internet of services (the third iteration, MISE 3.0, has been on-going since 2011).

The key scientific points of this research work are distributed among these three levels.

- *Business level*: (i) the use of a generic metamodel of collaboration (dedicated to collaborative situations) to formalize the situation, (ii) the ontology-based semantic reconciliation during the characterization step, and (iii) the automated deduction of a relevant collaborative behaviour in BPMN.
- *Technical level*: (i) the semantic and syntactic reconciliation between business elements and technical elements (data vs. information, activity vs. service, process vs. workflow) and (ii) the automated deployment on a service-oriented middleware of an operational mediation information system.
- *Agility level*: (i) the automated exploitation of incoming data to update situational models (based on the metamodel of collaboration), (ii) the detection of an adaptation necessity and the characterization of this necessity for adaptation.

The contributions of the MISE project to these key scientific points are mainly presented in the third, fourth and fifth sections of this article (presenting the business

---

[3] Enterprise Service Bus is a technical solution for a middleware respecting SOA principles

level, technical level and agility level of MISE 2.0). The first section of this article provides an overview of existing research works through a literature review, while the second section presents the whole MISE 2.0 approach. The final section concerns our conclusions about MISE 2.0 and the perspectives for MISE 3.0.

**1 Literature review**

Considering the global structure of the MISE approach (three levels), this section presents results and research works in five parts, dealing with, first, the legitimization of this structure, then with business issues, technical issues and agility issues (*i.e.* covering the three levels), and finally with the scientific positioning of these research works.

*1.1 Preliminary point: collaboration architectures*

The European Interoperability Framework (EIF), presented in [11] and discussed in [12], has been defined by the European Commission and by member states of the European Union. It can be considered as a relevant reference to structure the global architecture of the presented research work (in that it addresses the need for efficient information exchange). The three following dimensions may be extracted:

- Organization Interoperability: this level concerns business goals and collaborative processes and is mainly based on Business Process Management (BPM).
- Semantic Interoperability: This level aims at achieving meaningful exchanges and concerns the gap between data and information. It is mainly dedicated to building a shared knowledge base.
- Technical Interoperability: This level deals with connectivity and ensures that organizations get physically connected. This is abusively often the common meaning given to the term interoperability (because it is currently the most accessible level).

Considering the MISE structure (*business* level, *technical* level and *agility* level), one can notice that the first two levels are very close, respectively, to *organization interoperability* and *technical interoperability*, as defined in the EIF. Furthermore, *semantic interoperability*, which is clearly identified in EIF as an independent level, is actually split between both the *business* and *technical* levels of MISE:

Knowledge gathering and deduction through ontologies at the *business* level clearly requires ensuring "business semantic reconciliation" between partners (and also with existing knowledge bases). Furthermore, transforming business processes into technical workflows at the *technical* level (including data alignment and service discovery) also requires ensuring "technical semantic reconciliation".

The so-called *agility* level of MISE directly refers to the dynamicity of collaborations, which is not considered by EIF. Consequently, the MISE structure (*business*, *technical* and *agility* levels) is compliant with the EIF structure (*organization*, *semantic* and *technical* interoperability). The following part of this section explores these three levels of the MISE structure.

## 1.2 Literature review: business level

As this first level concerns the modelling of collaborative business processes, the main target of this subsection is research works and results for *business process emergence* (in a collaborative context). The associated scientific need concerns the support provided for collaborative business process design. Business Process Management (BPM) is consequently the appropriate scientific field to explore. However, BPM covers the whole lifecycle of business processes (as detailed in [13]) but this subsection will focus on the *design* and *modelling* phases (while the rest of this article also deals with the *execution*, *monitoring* and *re-engineering* phases). Actually, both these phases are relevant ones for the focus on the scientific need at the business level.

According to [14], a business process is "a series or network of value-added activities, performed by their relevant roles or collaborators, to purposefully achieve the common business goal", which is perfectly coherent with [15]. In the world of BPM, many different process modelling notations and tools have been proposed (e.g. IDEF Suite, BPMN, ARIS, UML, Structured Analysis and Design Technique, Petri Nets, Object Oriented Modelling, CIMOSA, IEM approach) and studied [16]. Their functionalities and characteristics vary, which can lead to misunderstanding and failure. Furthermore, executable languages used to implement the models (e.g. BPEL or classic programming languages) are also diverse. These issues are similar to those identified in the Model-Driven Software Development (MDSD) concept [17], which is a specialization of MDA. In [18], a summary of the software and tools used to describe business processes in a sample of companies is presented (see Table 1). A worldwide survey of major public companies has been conducted to elicit the requirements, which are grounded in the nature of processes and the usage of software. The analysis of 127 responses indicates that human-oriented process modelling languages and BPM tools, including BPM tools with software integration capabilities, are most urgently required. Obviously, many companies combine text and some modelling languages (55.9%), but tables are also widespread (31.5%). Among the languages, BPMN dominates, followed by the Unified Modelling Language (UML) and Event-driven Process Chains (EPC).

Table 1.  Current documentation of processes (N=127 and Na=3) [18].

There are numerous valuable research works in the business process management field, beside these well known modelling languages and modelling architecture, for example situation calculus [19], operational process formalization [20] or business process abstraction [21]. These research works focus on question of business process management and could give some clues to solving problems encountered in the MISE 2.0 business level design.

One very interesting development concerns SoaML [22], which provides a metamodel and a UML profile for designing services in a service-oriented architecture. The provided UML profile includes the following new modelling capabilities:

- The ability to provide a complete description of services (fulfilled requirements, dependencies, capabilities, protocols, providers and users)
- The ability to define policies for using and providing services
- The ability to link services with upstream models (to describe the needs for services) and downstream models (to describe the use of services)

SoaML is a very strong framework to describe both the *business* and *technical* points of view of services. As a UML profile, it is easy to use in a modelling perspective.

From the MISE point of view, the SoaML might be analysed according to the following statements: It is a very powerful description framework, able to ensure the path from service-oriented modelling of business systems (static and dynamic) to service-oriented orchestration / choreography of these business systems. However, some complementary comments might be stated: (i) SoaML requires the user to fully design the process models (no emergence of behaviour), (ii) SoaML also requires that services be modelled from scratch (for instance no extraction from WSDL files or SA-WSDL files) and (iii) SoaML provides a full service-oriented modelling framework but no execution framework (consequently, SoaML provides the tools to manage system agility but does not support it directly).

Finally, SoaML is a very rigorous and exploitable framework for service-oriented design of system. From the point of view of the service-oriented system lifecycle, the covered perimeter is not exhaustive but it is compliant with most of the expectations (from process modelling to workflow deployment).

Situation calculus can be applied to formally specify and analyse business processes by considering the intuitive mapping from an activity in a process to an action in the situation calculus domain [23]. Situation calculus can be briefly summarized as two tasks (i) automatically selecting business activities to achieve business goals and (ii) verifying that the selected functions complete the goal successfully and correctly. The first task, Automatic Service Composition, is consequently very relevant to this article.

Automatic Service Composition is dedicated to composing services automatically from the existing isolated services inside or outside an enterprise. To enable this automatic composition, formal descriptions of services are a prerequisite. A formal service description refers to specifying services by employing formal methods, usually mathematical logic. With such a formal specification, services are described precisely and unambiguously. Logical reasoning can be performed, which enables the automatic composition. The main drawback of this approach concerns the need for this formal description of services. For technical services (computerized ones), such a formal description is not a problem in itself, and is even a very common feature (WSDL, etc.), but with regard to business services, this is really an obstacle.

## 1.3 Literature review: technical level

The technical level deals mainly with the reconciliation of business processes and executable workflows. There are clearly three reconciliation issues, which, once solved, may allow business processes to be connected to technical workflows in a relevant manner: the *informational* issue (how to select data that faithfully represents business information?), the *functional* issue (how to ensure matching between business activities and technical services?) and the *behavioural* issue (how to obtain workflows from business processes?). Furthermore, these issues deal with many-to-many considerations (e.g. a set of $n$ activities may be ensured by a set of $m$ technical services).

Consequently, this subsection should address these three issues. However, because (i) *informational* reconciliation strongly depends on *functional* reconciliation (business activities deal with information as inputs or outputs and technical services deal with data as inputs or outputs) and (ii) *behavioural* reconciliation mainly depends on the ability to transfer the structure of business processes into the structure of technical workflows once the *functional* reconciliation has been done, the central issue is the *functional* one of replacing business activities (in business processes) with relevant technical services (in workflows), even in a many-to-many manner. Furthermore, this research work considers that business activities are the functional steps of business processes while technical services are the computed steps of workflows. The question is mainly to define which *human function*, *automated function* or *IT function* should be used (even in a many-to-many manner) to ensure a specific business activity. Besides, if an *IT function* could be easy to integrate into a workflow as a technical service, *human* and *automated functions* require interfaces (as technical services) to be integrated into any workflow. This subsection is consequently especially focused on service discovery as the main domain to explore for *functional* reconciliation [24].

In the literature, there are three main approaches dedicated to ensuring service discovery from a *functional* viewpoint: (i) syntactic approaches based on names, words and vocabularies, (ii) semantic approaches based on concepts and meanings and (iii) hybrid approaches, which combine both the previous ones. There are several existing frameworks and tools based on these approaches.

WSMX (Web Service Execution Environment), presented in [25], is supported by the ESSI[4] cluster. This framework aims at exploiting the semantic description of the web services provided, by using WSMO (Web Service Modelling Ontology) to manage web services discovery. WSMX uses the concepts of prerequisites, inputs, and outputs to sort web services. Furthermore, WSMX also takes into account QoS (Quality of Service) to refine the obtained sorting. The main drawback concerns the fact that it is only a one-to-one reconciliation between business activities and technical services, while the granularity of technical services is, in general, largely thinner than that of business activities.

SUPER is a European funded Integrated Project (FP6), which aims at managing the whole lifecycle of collaborative processes (from BPMN modelling to workflow execution). As presented in [26], this project is based on IRS-III (Internet Reasoning Service) and WSMX. SUPER uses SBPM (Semantic BPM) for a semantic description of processes, which includes:

- The description of the partners' processes in a process repository.
- The complete description of partners' information systems in an ontology.
- Domain specific knowledge (constraints, business rules).
- Modelling of data for information reconciliation.

Furthermore, SUPER provides service composition, which allows one-to-many reconciliation (if none of the technical services satisfies the requirements alone).

---

[4]ESSI cluster is a scientific initiative of the Semantic Technology Institute (STI) of Innsbruck http://www.essi-cluster.org

According to MISE requirements, the main drawback of these research works concerns the fact that SUPER does not use SAWSDL, which is currently the major standard for semantic description of web services.

FUSION [27] is very similar to SUPER (also a FP6 European project), however, although it deals with SAWSDL, it is not able to manage service composition (limited to one-to-one reconciliation).

SOA4All [28] is also a European funded project (FP7/IST), supported by the European platform NESSI[5], which aims at providing a suite of tools dedicated to supporting the use of web services. The semantic aspects of web services are described using WSMO-Lite (a light version of WSMO), while the semantic aspects of business processes are described using LPML (Lightweight Process Modelling Language). The reconciliation engine DTCE (Design Time Composition Environment) uses these elements to define executable workflows [29]. Furthermore, DTCE supports late binding. The main drawback of the SOA4All project is that, just like the SUPER project, SAWSDL is not supported.

There are also a lot of other research works concerning this subject (IRS-III, METEOR-S, etc.) and the following table presents a summary of their main functionalities.

Table 2. Synthesis of service reconciliation frameworks and tools.

## 1.4 Literature review: agility level

Agility is a crucial concept in a collaborative situation [30]. [31], [32] and [33] draw the line between this concept and reactivity, flexibility and adaptability. In [34], agility is

---

[5] Networked European Software and Service Initiative. http://www.nessi-europe.com

defined as the ability of a system to be *flexible*, in a *reactive* and *efficient* manner (cf. the *house of agility*). There are three main aspects to this vision: the system must be able to change its structure (*flexibility*) according to a relevant analysis of the situation and its requirements (*efficiency*) and this should be done in a hurry (*reactiveness*). In the context of the MISE project, these three facets of *agility* have also been considered according to two orders: first order represents the main components of agility while second order concerns the features of these main components. Consequently, agility has been defined, on first order, as the capacity of a system to (i) *detect* any (potentially unexpected) situation that requires the system to change and (ii) *adapt* its global structure/behaviour to that situation. Regarding second order, two other attributes may be considered: first, the *dynamicity* of agility might be crucial (performing *detection* and *adaptation* too slowly may disrupt agility) and second, the *relevance* of the *detection* and *adaptation* may also be critical (wrong *detection* and *adaptation* could be fatal for the significance of agility). Consequently, this vision may be simply and roughly formulated as:

**agility = (detection + adaptation)** × *(reactiveness + efficiency)*.

Such a formula, although not scientific at all, is a structuring scheme that allows the study of agility to be partitioned according to these three properties. Finally, *detection* and *adaptation* may be considered as the main attributes of *agility* while *reactiveness* and *efficiency* are the attributes of *detection* and *adaptation* (second order). Considering that *reactiveness* and *efficiency* are from a second order and specifically describe both *detection* and *adaptation*, the following will mainly focus on two precise scientific needs: first, the ability of a system to *detect* any relevant change on a defined space; and second, the ability of a system to *adapt* its own static and dynamic properties to satisfy the new requirements that have emerged from the change in the considered space.

Given that agility must be performed on a technical architecture (inherited from the technical level), the notions of *detection* and *adaptation* have to be restricted to a service-oriented domain.

As regards the "detection" aspect of agility (especially regarding information systems), it is definitely dependent on the ability to gather data about the current collaborative situation [35]. Actually, the technical architecture for MISE is SOA-based and consequently, this objective of data gathering invites the consideration of EDA (event-driven architecture) as a relevant approach. [36] presents a synthesis of this technology. EDA is considered as mediation architecture in which some components are event-driven and which uses events to communicate. Mainly, any kind of message sent by any component of an EDA-based system has to be considered as an event. The specificity of events [37] is that they are published in an "event market place" where they can be computed and forwarded to systems that are subscribers to specific events (in a type-based or content-based mechanism). This is the "publish/subscribe" mechanism. Furthermore, tools such as CEP (Complex Event Processing) may combine received events, according to pre-established event patterns, to publish new events [38]. Considering that events could be of any type, the use of this kind of tool is particularly appropriate, given that rough events cannot be directly inserted in the formal knowledge managed by the system. Combination, aggregation and patterns can be used to help to transform incoming events into formal instances. These formal instances can then be used to detect any changes.

Considering the "adaptation" aspect of agility, [39] presents a synthesis on collaborative process flexibility. The proposed taxonomy contains four types of adaptation approach:

- Adaptation by design: this concerns the design of very "hairy" processes, including a lot of alternatives, which allow the most appropriate path to be selected at run-time. This first approach implies exhaustively enumerating all the potential options that could be relevant during run-time, which is a difficult task (first drawback). Furthermore, this approach also requires precise description of a lot of paths that will never be considered during run-time (second drawback). Risk management, in its classic designation, is generally based on this approach [40].

- Adaptation by deviation: this approach aims at providing flexibility when it is required in run-time by allowing the modification of the sequence of activities without changing them. Tasks may be restarted, cancelled, avoided or re-organised.

- Adaptation by underspecification: in this approach, activities and processes are partially defined in design-time and will be completed in run-time [41]. Abstract tasks and processes (that are not completely described) are used to structure the whole behaviour, but during run-time, these abstract elements have to be specified in detail (only if they are relevant in run-time). Two types of underspecification may be used: (i) late choice of elements, *i.e.* late-binding [42], or (ii) late design of elements, *i.e.* late-modelling [43].

- Adaptation by change: In this fourth approach, the definition of the process may be modified during run-time, in exactly the same way as adaptation by deviation, but this approach is more powerful: tasks may be inserted and processes can be totally re-engineered.

The following figure presents this agility framework.

Figure 2. Different types of collaborative process agility [39].

## 1.5 Scientific positioning

Considering the three levels of MISE and the preceding literature review, the following scientific stakes have been identified. As a result, some research choices have been made, bringing in turn their own stakes.

First, at the business level, the main scientific stakes concern the gathering and transformation of informal knowledge into an exploitable model. To drive this BPM approach, the choice was made to use metamodels, ontologies and deduction rules (*i.e.* some familiar tools for knowledge management). The coverage of the domain of collaborative situations (especially by metamodels and ontologies) is the first major issue resulting from this choice. The relevance of the deduced behavioural scheme (collaborative process), based on the situation and the partners' capabilities and in the SOA context of deployment, is certainly the second major issue at stake in the choices made at this business level. Obviously, the resulting models might have used the SoaML framework [22] to formalize the collaborative behaviour in a SOA compliant way. However, the deduction process was not supported by SoaML and is clearly a contribution of MISE, outside SoaML perimeter.

Secondly, at the technical level, the main scientific stakes concern SOA-governance, orchestration and choreography of web-services. To meet these needs, the choice was made to use a hybrid approach (both syntactic and semantic) based on ontologies and semantic annotation features (for BPMN and WSDL). Service discovery (in a many-to-many manner) from collaborative business behaviour is the first issue associated with these choices. Furthermore, the "on-the-fly" reconciliation of information and data is the second issue at stake. Again, the SoaML framework could have been used to describe orchestrable workflows, however, the business to technology reconciliation is a specific feature of MISE that is not provided by SoaML (the business to technology projection is a human task in SoaML).

Third, at the agility level, the main scientific stake concerns the supervision of the collaborative situation and the exploitation of the knowledge generated by this supervision. There is a strong complementary issue here, in the fact that the supervised system (the collaborative situation) is initially an unknown system (in terms of content, boundaries and dynamic aspects). The choice has been made to use EDA and CEP to supervise the situation and to formalize the gathered knowledge according to metamodels. There are two main issues associated with this choice: (i) the interpretation of a various amount of event types (potentially unknown) to significantly update models and (ii) the relevant analysis of updated models to deduce the appropriate adaptation measures.

## 2   Overview of MISE 2.0

MISE 2.0 is the second iteration of the MISE project. The first iteration ran from 2004 to 2010 (including three doctoral works and the two funded national projects ISyCri and JOnES). The second iteration started in 2009 and ends in 2013. MISE 2.0 includes four doctoral works and three funded projects PLAY (European), SocEDA and ISTA3 (national). The third iteration MISE 3.0 started in 2011 and is currently on-going. It is also based on several doctoral works (currently five) and on funded projects (currently three: SIM-PeTra, OpenPaaS and DRIVER).

### *2.1 General principle of the MISE project*

The MISE project is dedicated to supporting collaborative situations. It can be extracted from [44] (and also from [8]) that there are four levels of collaborative situations: (i) simple exchange of information, (ii) actual connection of services and applications, (iii) definition of collaborative processes, (iv) full integration of partners (*i.e.* the same as the third level without any specific effort from partners). A parallel can be established between these collaboration levels and the three layers of organizations (respectively ISs), *i.e.* information (respectively data), activities (respectively applications) and processes (respectively workflows) presented in [10] (the fourth collaborative level can be considered as a refinement of the third one). The objective of MISE 2.0 to support any collaborative situation implies necessarily to deal with these three layers of organizations (respectively IS). These considerations explain and justify the main orientation of MISE 2.0 on collaborative processes.

Furthermore, service-oriented technologies provide an environment able to deal with workflow orchestration (including exchange of data, connection of services and evolution of workflows). Consequently, such an environment is a very appropriate candidate to meet the requirements inherited from the three layers of organisations (respectively ISs) and consequently to provide MISE 2.0 with the adequate platform to support any type of collaborative situations.

The overall MISE design approach might be seen as a dive into abstraction layers based on model-driven engineering [45]. The general principle of the MISE approach (whatever the iteration considered) is structured according to four steps (two at the business level and two at the technical level):

(1) Design of collaboration model: this level concerns the gathering of knowledge about the considered collaborative situation to instantiate concepts of the so-called collaborative metamodel (concerning mainly the *environment* of the collaboration, the *objectives* of the collaboration, and the *partners* and *services* of the collaboration).

(2) Deduction of collaborative behaviour model: the second step deals with the automated deduction of collaborative processes, based on the knowledge collected at the previous level. Schematically, the aim is to select and organize partners' *services* according to the *objectives* and *environment* of the collaboration.

(3) Design of collaborative workflows: the previously deduced business behaviour (processes) is translated into a technical behaviour (workflows) to be implemented. The goal is mainly to match services with activities and data with information.

(4) Deployment and orchestration of the MIS: the previously obtained workflows are integrated in a workflow engine to be executed on an ESB. All available web-services of the partners are connected on the same ESB (in case of necessity, specific interfaces are also deployed to connect other service or even human tasks). The collaborative behaviour is consequently performed on this middleware among partners' services.

Furthermore, these four steps are used in an agile framework, which deals with *detection of evolution* and *adaptation of behaviour*. The agility of the MIS is based on event analysis (according to the received event, is the situation in line with what is expected?) and on behaviour adaptation (by re-invoking step 1, step 2, step 3 or step 4, depending on the nature of the event analysis). From a technical point of view, the MISE project is based on a Service-Oriented-Architecture (SOA) paradigm and MISE tools are also deployed as web services on the same ESB as the partners' web services.

Figure 3. Overall structure of the MISE project.

Even if there are some differences and specific features, each of the three iterations of the MISE project is structured according to the four steps presented above, and the associated agile framework. Furthermore, from a technical point of view, these iterations are all centred on SOA principles and on web services.

### 2.2 *Specific improvements of MISE 2.0*

Considering MISE 1.0 as a foundation work, MISE 2.0 aims at reusing the results obtained and adding some new features. However, there were several drawbacks with the first version of MISE. The most important ones are the following:

- The use of domain-specific metamodels does not allow the approach to be relevant for any kind of collaborative situation (furthermore, the associated knowledge bases cannot be used conjointly).
- Deducing one single collaborative process is not very relevant because most organizations are structured according to a typology of processes (e.g. *decisional*, *operational* and *support* as recommended by ISO 9000-2001 [46]).
- The transition from *business processes* (embedding business activities and information) to *technical workflows* (concerning technical services and data) is too manual (automated workflow generation but manual selection).
- The adaptation functionality is assumed by the service-oriented structure (recall of design-time services to re-define the collaborative behaviour) but the detection functionality is fully manual, based on human analysis of reports.

To deal with these weak points, MISE 2.0 is managed according to the following main principles. First, one single metamodel (representative of collaborative situations) has been defined [47]. This metamodel, the instances of the associated ontology (*i.e.* the ontology structured according to this metamodel) and associated deduction rules (defined from concepts of the considered metamodel and dedicated to dealing with instances of the associated ontology) can thus be used in any collaborative situation. This structural improvement reduces the impact of the first listed drawback. In addition,

MISE 2.0 uses an objective typology to deduce a complete collaborative process cartography including several processes, which are typed as *decisional*, *operational* and *support* processes. This point tackles the second drawback. Besides, semantic reconciliation mechanisms have been injected (as described in [48]) to deal with the transition from business processes to technical workflow (*i.e.* the third drawback of the previous list). This improvement uses semantic annotations of business activities on the one hand, and of technical services on the other hand, to select the most appropriate subset of technical services to engender the behaviour described by the considered business activities. Based on semantic annotations of information, these research results also provide on-the-fly data translation to assume correct orchestration of the selected technical services (this tackles the third weak point). Finally, an event-driven architecture (including a CEP tool) is added to the service-oriented structure of the MIS. This improved technological platform provides two main points of interest. The first one concerns the choreography of multi-processes. Deducing a collaborative process cartography implies the ability to orchestrate each workflow, but also to manage the coordination of these workflows. Workflow orchestration is assumed by the SOA structure while the coordination of several workflows is assumed by the EDA structure (through choreography). The second point of interest concerns the detection aspect of agility. Services (but also other devices or sensors) are able to send events. These events might be used by the system to detect any unexpected situation. This diagnosis mechanism is a solution to the fourth identified drawback [49]. The following table summarizes the specificities of MISE 2.0 (compared with MISE 1.0).

MISE 1.0 and MISE 2.0 are associated with some concrete application fields. For instance, the *ISyCri* project concerns MISE 1.0 in a crisis management context [50], while the *ISTA3* project concerns MISE 2.0 in a manufacturing field [51].

Table 3. Specificities of MISE 2.0, compared with MISE 1.0.

## 3 Business level of MISE 2.0

For MISE 2.0 abstract level design, the main objective is to build the collaborative process cartography. The collaborative process of MISE 1.0 [9] was a "mixed" process, which covered the information involved in strategy, operation and support knowledge [46]. This kind of collaborative process is very difficult to understand and execute. The collaborative process concerns different levels of users (managers, workers from operating units or warehouses, and so on). These users own different knowledge, which makes them able to interact efficiently with parts of the collaborative processes. Consequently, it is better to build several specifically engineered collaborative processes, which present different parts of the "mixed" collaborative process. All of these collaborative processes should be managed and presented by a main process, which is the top-level of the collaborative process cartography. The collaborative process cartography classifies processes as strategy, operation or support processes [46]. The collaborative process cartography presents the process as one main top-level process (with the information of classification) and several low-level sub processes. The use of this three level classification may be questionable, however, it has been considered as the best solution to structure the process cartography.

To build the collaborative process cartography, relevant collaborative knowledge about the situation and the target network should be gathered and transferred. The principles are (i) to gather the essential and minimum initial collaborative knowledge (e.g. *partners*, collaborative *objectives* and shared *functions*) in the mode of a model, (ii) to deduce the missing knowledge with the help of an ontology/metamodel with transformation rules and (iii) to complete the collaborative process cartography with the deduced knowledge by means of algorithms. Based on the above principles, in a collaborative situation, the partners come together with their shared functions to achieve the objectives of the collaboration. The shared functions and the collaborative objectives could be seen as the initial collaborative knowledge. The goal of the MISE 2.0 abstract level is to select the shared functions and build the collaborative process cartography, which is made up of a main process and several sub processes of strategy, operation and support. These collaborative processes (i) provide the sequence of shared business functions to follow, (ii) embed strategy, operation and support processes for different levels of users, and (iii) are dedicated to achieving the collaborative objectives.

The "in" and "out" knowledge of the MISE 2.0 abstract level can be summarized in Fig. 4:

Figure 4. Collaborative network model, function model and collaborative process model.

The collaborative network model collects the initial collaborative knowledge, which includes the collaborative *context* and the collaborative *objectives*. The function model is necessary to describe the *capabilities* of the partners. Furthermore, each function must contain a semantic annotation for further transformation from business processes to executable workflows. Then the collaborative process model is defined to represent the *behaviour* of the collaboration.

MISE 2.0 abstract level can be summarized as four main phases. All the details of MISE 2.0 abstract level are described in [52].

(1) Knowledge Gathering: the knowledge in this phase covers the target collaborative situation. In [9], initial knowledge is structured according to concepts like *collaborative network, partners* and *common goal*. In [50], the shared *functions* of partners are added to the initial knowledge.

In MISE 2.0, the above two results are combined together and improved. The collaborative network model and function model represent and define the initial collaborative situation. The collaborative network model does not only collect the collaborative network, partners and partner relations but also the collaborative sub networks, and collaborative objectives. The function model represents the information concerning shared partner functions and the associated input/output messages.

(2) Knowledge Transfer: in this phase, the collaborative ontology and transformation rules (defined in the MISE project for generic collaborative situations) are used to create *mediation instances* in the collaborative ontology from *collaboration instances*. Actually, the deduction of mediation instances implies several steps that are interdependent (in a waterfall structure). Consequently, there are five groups of transformation rules (as explained in [52]): the first group is dedicated to creating the *Mediator*. The second group is dedicated to creating *Mediator Relationship*s. The third concerns the creation of *Generated Mediator Functions*. Group four links the *Generated Mediator Functions* to the *Mediator*. Finally, the last group is dedicated to creating *Inter Mediator Functions*. Tables 4 and 5 present the transformation rules of groups 1 and 2. With these transformation rules, mediation instances are deduced, but there is not enough knowledge yet for the extraction of a collaborative process.

Table 4. Example of rule in group 1.

Table 5. Example of rule in group 2.

(3) Knowledge Completion: this phase concerns the matching between *objectives* and *functions*. It is based on the selection of *business functions* to achieve *objectives* by linking the particular *functions* and *objectives* of the considered collaborative situation to the instances (of functions and objectives) of the collaborative ontology by using "same as" and "near by" relations.

(4) Knowledge Extraction: this last phase covers the structural design of collaborative processes by arranging business functions, through the deduction of sequences and gateways. In this phase, the specific deduction rules (generically defined in the MISE project to achieve the structural design of collaborative processes) are used to build the structure of the collaborative process cartography and of collaborative processes. First, the overall cartography of processes is deduced, based on the *strategic*, *operational* and *support* level. This cartography only contains "high level processes" and describes the main functions required by the collaborative situation (behavioural coverage).

Figure 5. Example of a deduced collaborative cartography of processes.

Second, each "high level process" is described in a second layer of BPMN diagrams in order to describe the precise sequence of atomic business activities:

Figure 6. Example of a deduced collaborative process ("deliver product") from the previous collaborative cartography of processes.

One of the main issues concerns the integration of gateways in the activity sequence. To complete the sequence and the gateway, the method of sequence deduction is developed.

To support the models, the collaborative ontology, the transformation rules and the methodologies, the Mediator modelling 2ool is designed and implemented as Software as a Service (SaaS). The tool mainly implements the following functions: (i) defining the collaborative network model and the function model, (ii) importing the instances of the collaborative ontology and helping to choose "same as" and "near by" instances for the defined objectives, functions and input/output messages, and (iii) transferring the defined models to the collaborative process cartography by implementing the transformation rules of the collaborative ontology.

In MISE 2.0, the collaborative process cartography is the output of the business level and the input of the technical level.

## 4   Technical level of MISE 2.0

Once the business process cartography has been designed, the aim is to generate the associated MIS. To this end, business requirements must be matched with the information system capabilities of the partners. The final objective is to generate executable workflows that fit the cartography.

### *4.1 Semantic enhancement of business and technical models*

Matching business models with technical ones requires workable links between them. To achieve these links, the matchmaking mechanism proposes the use of available semantic information from both models. However, to facilitate semantic matching between business activities and multiple semantic web-service (SWS) representations, a common semantic profile has been defined. This semantic profile is suitable for both business activities and technical services. It embeds the required information and is fillable by semantic models (associated to BPMN 2.0 business processes for business activities description or WSDL files for technical service descriptions). This common semantic profile aims to facilitate service matching, and is the matching pivot between business activities and technical services. It allows users to express the functional aspects of models to describe activity requirements and service capabilities. It also includes high-level semantic description of inputs and outputs. Technical details do not matter for service matchmaking: interoperability problems will be solved during message matchmaking, once real services are chosen, by focusing on XSD annotated schemas or other technical representations.

Whereas a lot of annotation mechanisms exist for web services (such as SAWSDL [53] or WSMO-Lite [54]), the recent BPMN 2.0 is still lacking a semantic standard. However, among its useful features, such as its high design range (from very high level processes to executable workflows), this recent modelling standard brings an XML representation and its extension mechanism. Therefore, a specific annotation mechanism has been proposed. It is called SA-BPMN 2.0 and is based on a previous semantic profile. Figure 7 represents the two new XML tags. While the first one, called *SemanticDetails*, allows users to describe the functional requirements of business activities, the second one, called *SemanticElements*, enables the description of associated messages.

Figure 7. Semantic annotation for BPMN 2.0 (SA-BPMN).

### *4.2 Matching methodology*

At a concrete level, the main goal is to generate a mediation information system based on business process cartography (from the solution layer) and focused on SOA principles. In order to execute abstract processes, the appropriate BPEL processes (Business Process Execution Language) must be generated, which can be executed on an orchestration engine. BPMN 2.0 specifications already suggest a BPMN to BPEL syntactic mapping. This mapping allows processes to be transformed from one meta-model to another, but it does not bring execution-needed information such as real service endpoints or exact exchanged messages. This "abstract to concrete" transformation involves taking into account both the granularity and conceptual differences. This model transformation uses both top-down and bottom-up approaches. It is made up of three steps.

Step 1 concerns the matching of business activities to technical services. Abstract processes are designed using business activities, which differ from technical services (granularity levels, semantic concepts, etc.). It involves an "n-to-m" matching and ontology reconciliation of concepts from different levels. The syntactic BPMN to BPEL transformation (cf. step 3) is already handled by the BPMN management library used. So the need is to focus on the business-to-technical-process transformation. Therefore, the choice was made to exploit both the defined BPMN 2.0 extension mechanism and the SWS representations. Both bring semantic description of functional capabilities/requirements, inputs and outputs, in both business and technical models. In order to provide reusability and acceptable performances, the process transformation is based on a pattern database populated with previous successful tries. The whole process transformation follows these steps:

1. For each business activity, existing patterns are searched in the database.
2. For uncovered activities, the semantic descriptions of these activities are studied along with the available web services (thanks to the level-independent semantic profiles). The matching mechanism is designed to handle both "1-to-1" and "n-to-m" matching to fit business requirements with technical services in spite of granularity differences.
3. The matchmaking result is then presented to the user to validate technical choices. If some activities are still uncovered, the user can choose to develop a new web service, find another partner who already owns it, or entrust the proposed library to generate GUI-based services, which handle expected messages but entrust the added-value treatment to an external (human) actor.

Step 2 is dedicated to enabling "on the fly" data transformation. The discovery of web services that fit the functional needs is not sufficient to generate executable processes and ensure good communication. Data interoperability between them should also be provided. Three steps have been defined and can be applied to each web service to generate transformation for its inputs:

1. A matching between input concepts and previous service outputs is made, using semantic matching.
2. By means of the ontology, every concept is discomposed into sub-concepts that are as low as possible.
3. Using the databases, syntactic transformation rules are defined (for instance from American date to British date) or mathematic transformation rules (for instance, from Celsius temperature to Fahrenheit temperature).

Steps 1 and 2 are complementary: to perform message transformation, technical information about inputs and outputs (I/O) are required, i.e. to know which technical service is used. Then, step 2 uses the results of step 1. If no transformation is possible, another service must be found, through the step 1 mechanism.

Step 3 uses the BPMN-to-BPEL library to create the final BPEL file from the overall BPMN structure of the business processes (initial process cartography), selected web services (from step 1) and data transformations (from step 2).

Each step is made up of an automated search for best solutions, using the hybrid approach, then a manual selection, completion or validation by the business user. The whole approach described in this part has already been implemented as a prototype. A collaboration definition platform was produced. This platform allows users to annotate models produced by the previous step (business level), then generate executable workflows thanks to semantic matchmaking mechanisms.

## 5  Agility level of MISE 2.0

Previous sections have presented how a mediation information system, based on a service-oriented architecture, may be deployed to ensure interoperability and collaboration for enterprise networks: collaborative processes are designed (from gathered knowledge regarding the target collaborative situation), and executed (as collaborative workflows obtained thanks to an hybrid reconciliation) among web-services of partners with the aim of giving operational support to the collaborative situation. However, this is a waterfall approach, which does not natively allow any agility (as defined in section 1.4): the MIS is deployed exactly as it is supposed to fit the initial collaborative situation model. However, the collaborative situation may evolve and change. This introduces constraints at the Run-time level. The operational dynamics of collaboration may be exposed to some unanticipated unknowns that can require an evolution of the MIS. As explained in [55] and refined in [56], there are three kinds of sources of adaptation:

- The evolution of the collaborative situation itself: the perceived characteristics of the collaboration (context, and/or partners, and/or objectives) are no longer the same and need new collaborative behaviour.

- The evolution of the collaborative network of partners: the main issue here would be the arrival or departure of partners, but there might also be a question of the availability of capabilities. Due to any circumstance (lack of resource, temporary problem or overestimation of capability), any partner may finally not be able to assume one activity that he was theoretically expected to ensure.

- The potential dysfunction of the execution of a service (leading to the interruption of a workflow): although the deduced business processes cover the expected objective, and even if the designed technical workflows invoke relevant web-services, there might still be a dysfunction in the execution of these services (or it might just be a question of silly choices). Such a dysfunction needs to be detected.

Now that we have identified the potential sources of problems that would require agility, let us present the whole architecture of the MISE 2.0 implementation, which should provide these agility features. First of all, the whole MISE 2.0 approach is supported in a concrete way by technical components that implement the different abstraction levels of the waterfall structure. Each component has been designed as a service to be deployed on an Enterprise Service Bus (ESB). The following picture shows this architecture used in Design-Time:

Figure 8. Technical architecture of the environment and tools of MISE 2.0.

(1) The workflow engine of the ESB runs the design-time workflow (*i.e.* the different steps of MISE). (2) A modelling service (model editor) is used to gather and formalize the knowledge about the target collaborative situation (as *objectives*, *context* and *partners* information). (3) The obtained model is integrated in a collaborative ontology (embedding a large amount of instances, especially the ones extracted from the MIT Process Handbook, [57]). (4) Deduction rules are executed on this ontology to select and structure adequate business activities (provided by the partners involved) as collaborative business processes. (5) The obtained model (process cartography) can be injected into a reconciliation service. (6) The hybrid reconciliation service applies semantic and syntactic mechanisms to transform the business processes into technical workflows (thanks to the reconciliation of activities/services and information/data). (7) The obtained workflow files (BPEL) can be injected into the workflow engine of an ESB. (8) Collaborative run-time behaviour is then defined between partners, which are able to provide web-services for their own contribution to the collaborative network.

This architecture is totally able to support the "adaptation" part of agility. Thanks to the use of an ESB (SOA), it is possible to re-invoke and re-start any service among these design-time services to re-deduce more appropriate collaborative workflows. Considering the previously presented sources of adaptation, it is easy to suggest that in the case of situational change, re-start should concern knowledge gathering (step 2), in the case of network change, re-start should concern deduction of collaborative processes (step 4) and in the case of dysfunction, re-start should concern service discovery (step 6).

Regarding the "detection" part of agility, MISE 2.0 added an event-driven-architecture (EDA) layer to this service-oriented architecture (SOA). The principle is to receive all the events (information) that are published by run-time services (*i.e.* monitoring events) on the one hand and all the events (information) that are sent by the "field" (*i.e.* reports, sensors measurements, and other data provided by the collaborative network itself) on the other hand. The initial model of the collaborative situation (the one used to generate the collaborative behaviour) is then duplicated to obtain two models: the *expected model* and the *field model*. Both these models are then updated by a CEP engine (complex event processing tool), which uses received events to insert, delete or modify instances of the *field model* (if the received event come from the field) and of the *expected model* (if the received event is a monitoring event). For instance, one temperature measurement received from a sensor will allow the CEP engine to update the temperature attribute of the instances concerned, while a status event coming from one service will allow the CEP engine to infer that one business activity is over and that the objective that this activity is supposed to ensure may be considered as achieved. Consequently, two pictures of the collaborative situation are maintained through two different ways. By comparing and measuring distance between both these models, it is possible to detect relevant divergence between the real situation (represented by the *field model*) and the expected situation (represented by the *expected model*). The following figure presents this mechanism:

Figure 9. Updating of models (field and expected) in time.

Furthermore, because this distance measurement concerns clouds of points (and not just points), it is possible to characterize that distance to obtain qualitative knowledge about the difference between the two models. This knowledge allows the system to detect whether the difference mainly concerns the situation itself, the network of partners or the execution of services. Then, it is possible to deduce what type of adaptation should be made (knowledge gathering, deduction of collaborative processes or service discovery). The following figure presents the agile framework of MISE 2.0:

Figure 10. Agile Run-Time framework of MISE 2.0.

In addition to the eight steps previously presented, there are also the following steps: (9) events emitted by monitoring (services) or by the field (devices) are received and sent to the CEP engine (which is in a cloud architecture). (10) These events are treated by the CEP engine to update *field* and *expected models*. (11) Both these models are compared to find out if adaptation is required (if distance is over a given threshold) and what adaptation is required (according to the nature of that measure of distance). This makes it possible to select which adaptation workflow should be used. (12) The workflow engine interrupts the run-time workflow execution to start the appropriate design-time workflow (which will produce the new relevant run-time workflow).

**Conclusion and perspectives**

The MISE project, through its two first iterations, provides a concrete way to implement the collaboration lifecycle: (i) characterization and deduction of collaborative behaviour, (ii) computerization of collaborative behaviour and deployment of a mediation information system, (iii) monitoring and supervision of the running collaborative situation for corrective deduction of collaborative behaviour (and back to ii). This lifecycle can be considered as an illustrative realization for connecting the *Internet of Knowledge* (point i), with the *Internet of Services* (point ii) and the *Internet of things* (point iii).

With regard to the scientific contribution of MISE 2.0, there are essentially three major points: first, the knowledge management performed in MISE 2.0 (business level) allows collaboration ontologies and unknown collaborative situations to be connected, thanks to the use of collaborative metamodels (describing the generic concepts of a collaborative situation). The generation of relevant collaborative behaviours, based on the ontologies, is mainly a consequence of formalizing this knowledge management.

Secondly, the transformation performed in MISE 2.0 (technical level) of business processes into technical workflows (taking into account many-to-many considerations and on-the-fly data translation) is the concrete achievement of one objective of BPM (the two other are *certification* and *analysis for the improvement of enterprise behaviour*). The semantic gap here is a huge obstacle to overcome and MISE 2.0 provides an effective way of dealing with this issue.

Lastly, the monitoring and control of the collaborative situation performed in MISE 2.0 (agility level) is also a strong contribution, merging design-time and run-time by using the specificities of the chosen technical architecture, as described in [58].

The originality of the whole approach lies mainly in the full and continuous coverage of the collaboration from the emergence of the collaborative situation to its steady state. Furthermore, the whole approach is based on web technology and very compliant with the existing ISs of potential partners.

As regards the outlook for the next iteration of MISE (MISE 3.0), and based on the results of MISE 2.0, the main orientations may be the following: on the business level, there are three main improvements that could be considered. First, the deduction of collaborative behaviour could be less deterministic: several options of collaborative behaviour could be deduced according to several priorities (price, time, efficiency, effectiveness, etc.). These alternatives could then be s(t)imulated to give a qualified panel of candidate collaborative behaviours. Secondly, the deduction of a set of performance indicators could come with the deduction of each model of collaborative behaviour. This would allow workflows to be monitored more efficiently during run-time. Thirdly, the use of the ISO classification of processes (strategic, operational and support processes) is really questionable as far as the classification of a process might change depending on the considered viewpoint. Consequently, the next iteration of MISE should definitely consider another way for structuring the process cartography. Especially, in the context of using an event-driven architecture to manage process choreography, the way to structure the process cartography could be driven by interactions and impacts between processes (and the concerned objectives).

On the technical level, there is one main improvement that could be considered. It concerns technical interoperability, as defined by EIF: in reality, some business activities might not be covered by existing technical services (for instance human tasks). However, workflows must be continuous. Consequently, interfaces need to be developed and deployed to fill these gaps.

With regard to the agility level, there is one main improvement that could be considered. This perspective mainly concerns the detection criteria. Currently, detection is based on significant divergences (addition, deletion or modification of instances). However, if performance evaluation were used, the decision to start the adaptation mechanism could be refined. Furthermore, this enhancement of the detection principle is perfectly in line with the integration of performance indicators (cf. perspectives at business level).

Table 1. Current documentation of processes (N=127 and $N_A$=3) [10].

| Answers | Count $C_i$ | Percentage ($C_i/\Sigma C_i$) | Percentage responses ($C_i/N$) |
|---|---|---|---|
| As text | 71 | 36.2% | 55.9% |
| As tables | 40 | 20.4% | 31.5% |
| As flow charts | 2 | 1.0% | 1.6% |
| With Language: | | | 21.3% |
|    BPMN | 27 | 13.8% | |
|    UML | 19 | 9.7% | 15.0% |
|    EPC | 16 | 8.2% | 12.6% |
|    BPEL | 5 | 2.6% | 3.9% |
|    IDEF | 4 | 2.0% | 3.1% |
| Other | 12 | 6.1% | 9.4% |
| Total ($\Sigma C_i$) | 196 | 100% | 154.3% |

Table 2. Synthesis of service reconciliation frameworks and tools.

| Framework | Language | Parameters* | Approach | binding |
|---|---|---|---|---|
| WSMX | WSMO | IOPE | Semantic | 1-1 |
| IRS-III | WSMO | IO | Semantic | 1-1 |
| WSMO-MX | WSMO | IOPE | Hybrid | - |
| OWLS-MX | OWL-S | IO | Hybrid | - |
| SAWSDL-MX | SAWSDL | IOOp | Hybrid | - |
| METEOR-S | SAWSDL | IOOp | Semantic | 1-1 |
| SUPER | WSMO | IOPE | Semantic | 1-n |
| FUSION | SAWSDL | IOOp | Semantic | 1-1 |
| SOA4All | WSMO-Lite | IOOp | Semantic | 1-n |
| DynamiCoS | Specific | IOPE | Semantic | 1-n |

*IO: Inputs/Outputs – PE : Prerequisites/Effects – Op: Operation

Table 3. Specificities of MISE 2.0, compared with MISE 1.0.

|  | **MISE 1.0** | **MISE 2.0** |
|---|---|---|
| **Collaborative situation model** | Domain-specific metamodels are defined, depending on considered business fields (crisis management, manufacturing context) | One generic metamodel dedicated to all types of collaboration is defined (including external layers, enclosing domain specific concepts) |
| **Collaborative behaviour model** | One single collaborative process is deduced from the gathered knowledge. | Decisional, Operational and Support processes deduced from the gathered knowledge. |
| **Collaborative workflow model** | After manual identification of technical services (or user-interfaces) that would assume identified business activities of the deduced collaborative process, the process is translated in BPEL language to be executable. | Automatic semantic reconciliation allows selection of subsets of technical services that will be invoked to assume business activities from a technical point of view. Furthermore, ontological tools ensure "on-the-fly" data conversion. |
| **Deployment and orchestration** | Design-time tools are deployed as web-services on the same ESB as partners' (run-time) web-services. A workflow engine is used to orchestrate the collaborative workflow (orchestration). | Design-time tools are deployed as web-services on the same ESB as partners' (run-time) web-services. A workflow engine is used to execute the collaborative cartography of workflows (choreography). |
| **Agility** | Detection is a manual task based on the way situation evolves. Once a need for adaptation is detected, design-time tools (model editor, process deducing tool, workflow translator) may be deliberately invoked to (re)define the collaborative behaviour appropriate for the "new" situation. | Detection is based on EDA. Sensors and services publish events (reports on the situation or on workflow progress) that can be used to update situational models. If the current model differs from the expected model, then adaptation must be started based on the same principle than MISE 1.0. |

Table 4. Example of rule in group 1

| **Group 1: Create *Mediator*** | |
|---|---|
| ***Sub Network→Mediator*** | |
| ∀Sub Network (X) (∀hasPartner(Sub Network (X), Partner ($X_1$)) ∧(∀hasPartner(Sub Network (X), Partner ($X_2$))∧…∧ (∀hasPartner(Sub Network (X), Partner ($X_n$))) → ∃Mediator (X) ∧∃hasMediator (Sub Network (X), Mediator (X)) | **(1)** |

Table 5. Example of rule in group 2

| **Group 2: Create *Mediator Relationship*** | |
|---|---|
| ***Strategy* and *Operation Objective*→*Main Function*→*Business Message*→*Order*** | |
| If ∀Strategy Objective ($X_1$) (∀generates (Strategy Objective ($X_1$), Main Function ($X_1$))) ∧ ∀Operation Objective ($X_2$) (∀generates (Operation Objective ($X_2$), Main Function ($X_2$))) <br> If ∀Main Function ($X_1$) (∀out (Main Function($X_1$), Business Message (m))) ∧ ∀Main Function ($X_2$) (∀in (Main Function($X_2$), Business Message (m))) <br> → ∃ Order (m)(hasMediatorRelationship (Mediator ($X_1$), Order (m))) <br> ∃ Order (m)(hasMediatorRelationship (Mediator ($X_2$), Order (m))) | **(2)** |

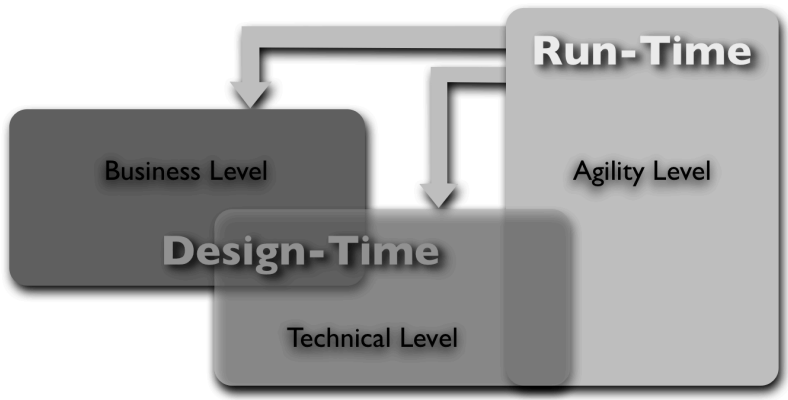

Figure 1. The MISE structure.

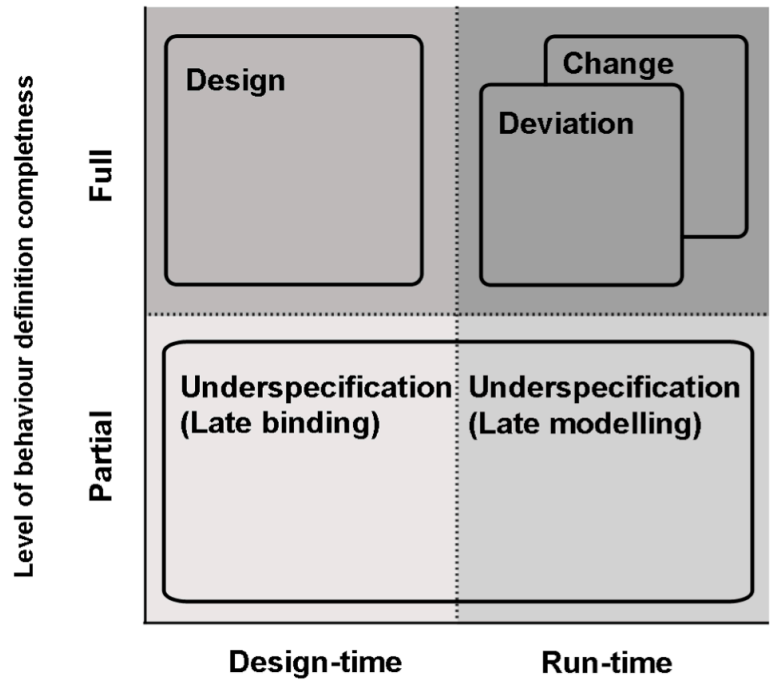

Figure 2. Different types of collaborative process agility [36].

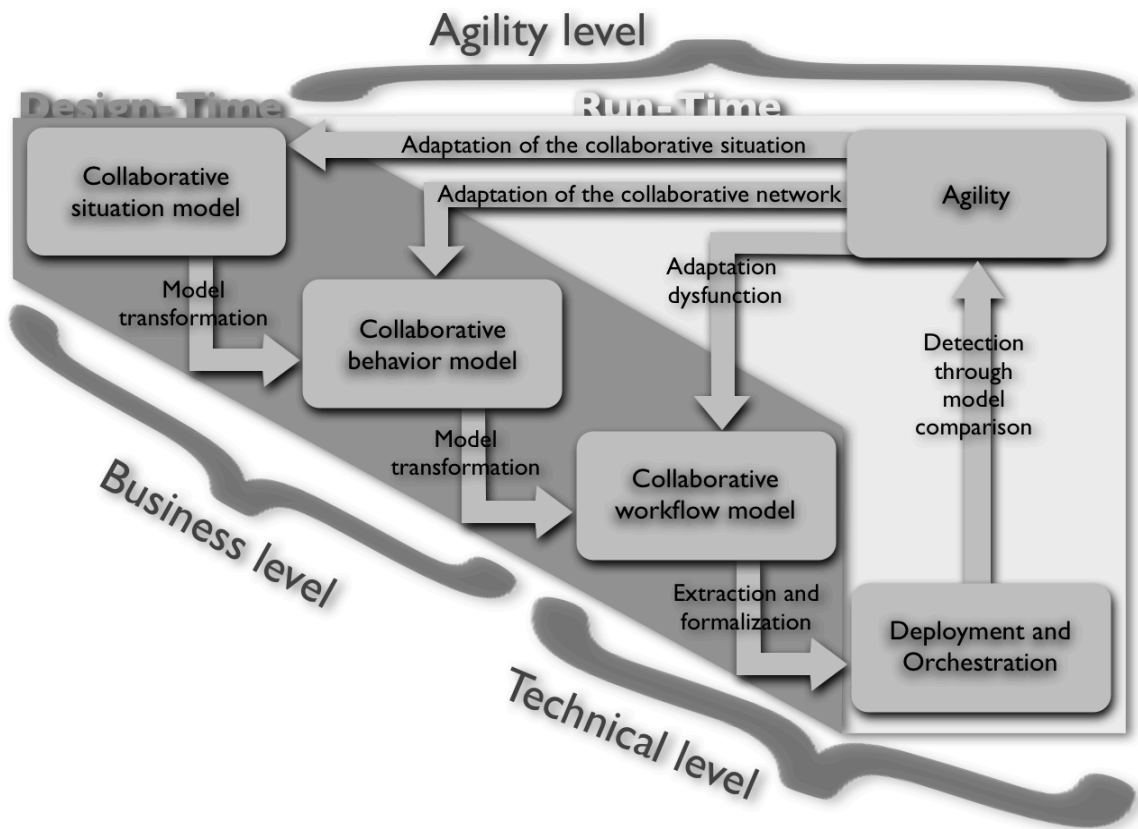

Figure 3. Overall structure of the MISE project.

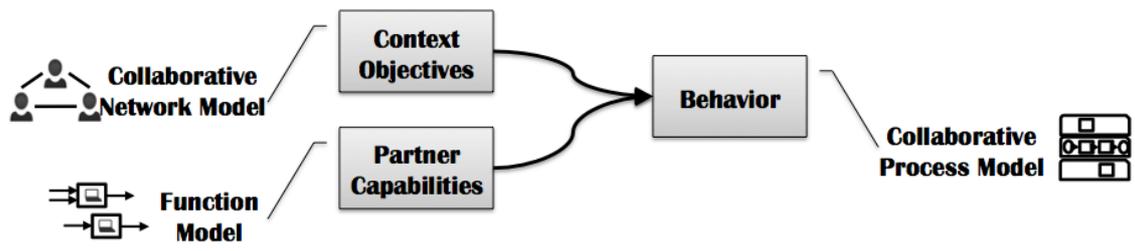

Figure 4. Collaborative network model, function model and collaborative process model.

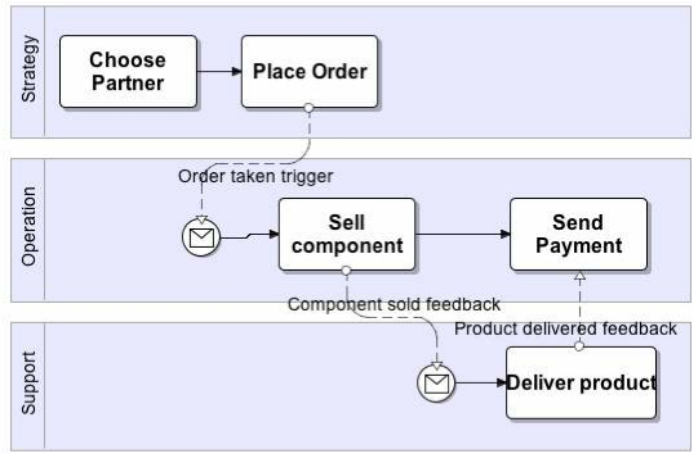

Figure 5. Example of a deduced collaborative cartography of processes.

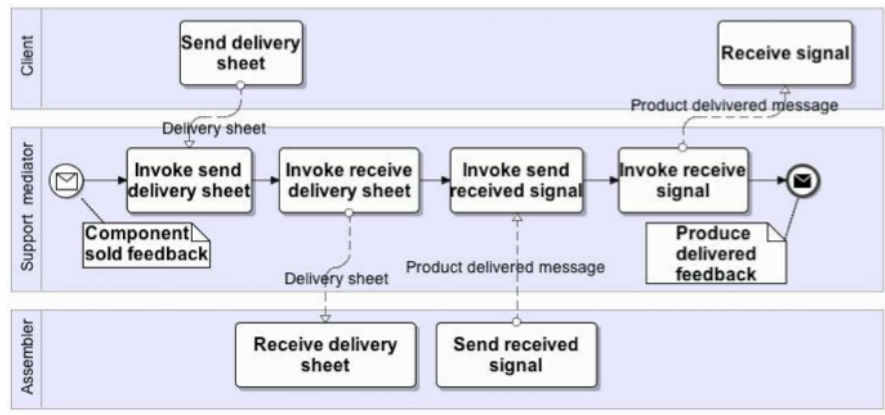

Figure 6. Example of a deduced collaborative process ("deliver product") from the previous collaborative cartography of processes.

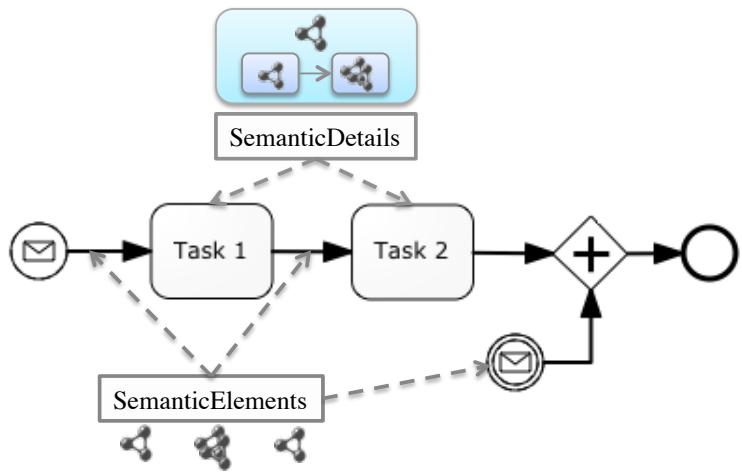

Figure 7. Semantic annotation for BPMN 2.0 (SA-BPMN).

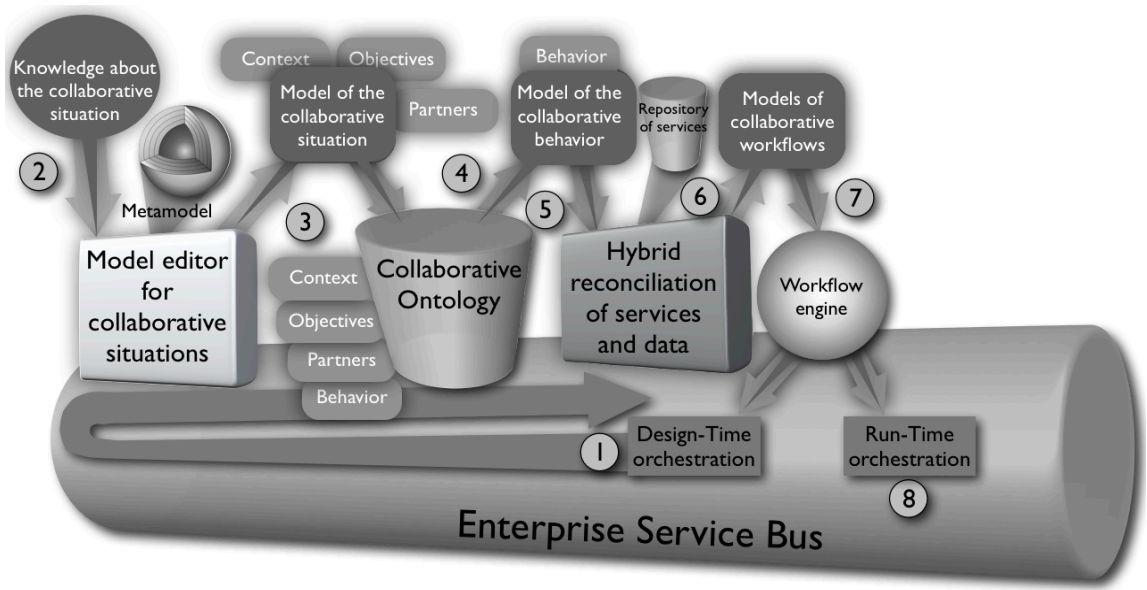

Figure 8. Technical architecture of the environment and tools of MISE 2.0.

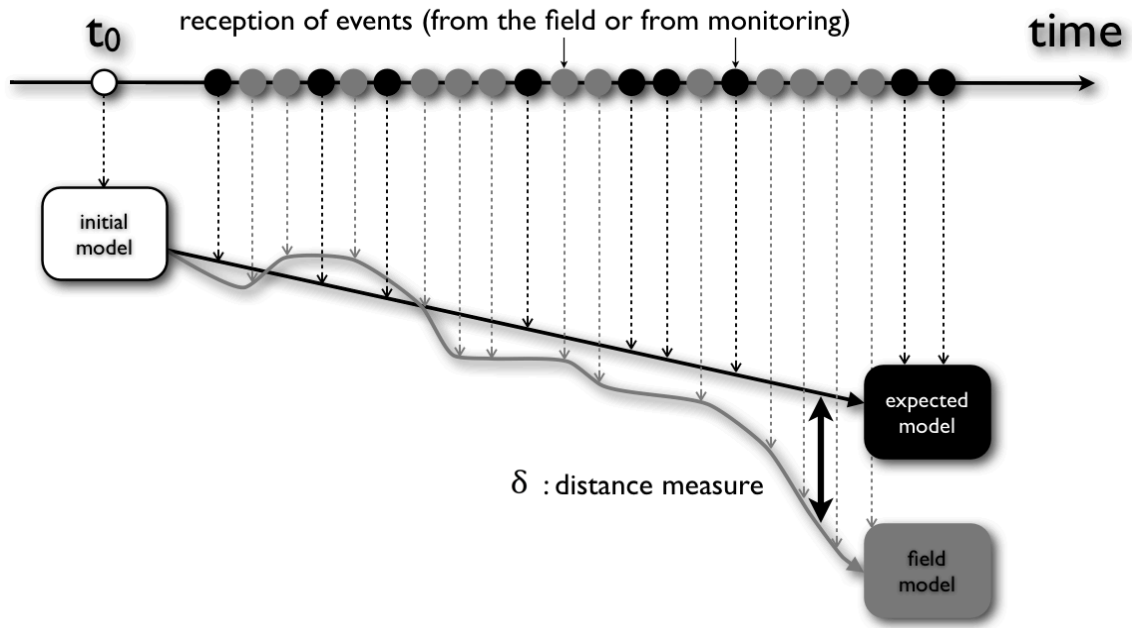

Figure 9. Updating of models (field and expected) in time.

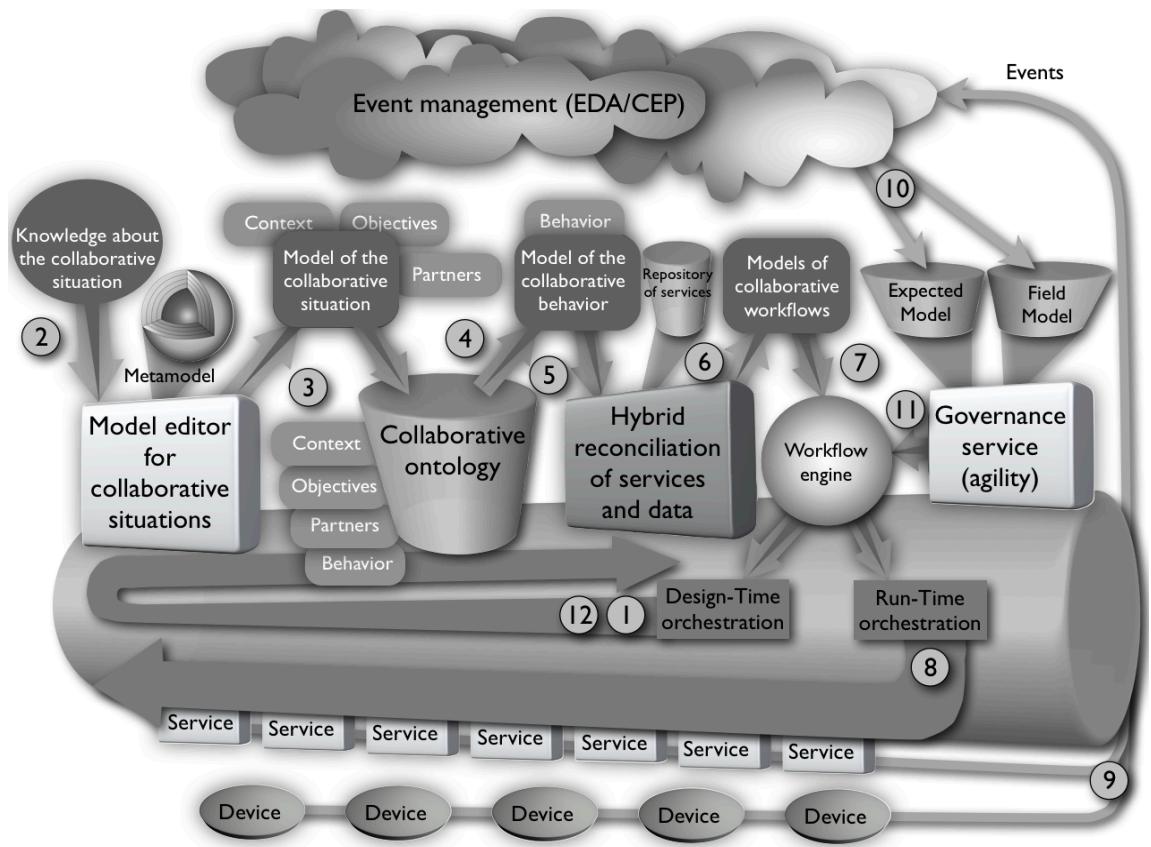

Figure 10. Agile Run-Time framework of MISE 2.0.